# A Conceptual Approach to Complex Model Management with Generalized Modelling Patterns and Evolutionary Identification

*Sergey V. Kovalchuk*[1], *Oleg G. Metsker*[1], *Anastasia A. Funkner*[1], *Ilia O. Kisliakovskii*[1], *Nikolay O. Nikitin*[1], *Anna V. Kalyuzhnaya*[1], *Danila A. Vaganov*[1,2], *Klavdiya O. Bochenina*[1]

[1]ITMO University, Saint Petersburg, Russia

[2]University of Amsterdam, Amsterdam, Netherlands

kovalchuk@corp.ifmo.ru, olegmetsker@gmail.com, funkner.anastasia@gmail.com, kisliakovskiii@yandex.ru, nikolay.o.nikitin@gmail.com, kalyuzhnaya.ann@gmail.com, vaganov@corp.ifmo.ru, kbochenina@corp.ifmo.ru

**Abstract.** Complex systems' modeling and simulation are powerful ways to investigate a multitude of natural phenomena providing extended knowledge on their structure and behavior. However, enhanced modeling and simulation require integration of various data and knowledge sources, models of various kinds (data-driven models, numerical models, simulation models, etc.), intelligent components in one composite solution. Growing complexity of such composite model leads to the need of specific approaches for management of such model. This need extends where the model itself becomes a complex system. One of the important aspects of complex model management is dealing with the uncertainty of various kinds (context, parametric, structural, input/output) to control the model. In the situation where a system being modeled, or modeling requirements change over time, specific methods and tools are needed to make modeling and application procedures (meta-modeling operations) in an automatic manner. To support automatic building and management of complex models we propose a general evolutionary computation approach which enables managing of complexity and uncertainty of various kinds. The approach is based on an evolutionary investigation of model phase space to identify the best model's structure and parameters. Examples of different areas (healthcare, hydrometeorology, social network analysis) were elaborated with the proposed approach and solutions.

## 1 Introduction

Today the area of modeling and simulation of complex systems evolves rapidly. A complex system [1] is usually characterized by a large number of elements, complex long-distance interaction between elements, and multi-scale variety. One of the results of the area's development is growing complexity of the models used for investigation of complex systems. As a result, contemporary model of a complex system could be easily characterized by the same features as a natural complex system.

Usually, a complexity of a model is considered in tight relation to a complexity of a modeling system. Nevertheless, in many cases, the complexity of a model does not mimic the complexity of a system under investigation (at least exactly). It leads to additional issues in managing a complex model during identification, calibration, data assimilation, verification, validation, and application. One of the core reason for these issues is the uncertainty of various kinds [2,3] applied on levels of system, data, and model. In addition, complexity is even more extended within multi-disciplinary models and models which incorporate additional complex or/and third-party sub-models. From the application point of view, complex models are often difficult to support and integrate with a practical solution because of a low level of automation and high modeling skills needed to support and adapt a model to the changing conditions.

On the other hand, recently evolutionary approaches are popular for solving various types of model-centered operations like model identification [4], equation-free methods [5], ensemble management [6], data assimilation [7], and others. Evolutionary computation (EC) provides the ability to implement automatic optimization and dynamic adaptation of the system within a complex state space. Still, most of the solutions are still tightly related to the application and modeling system.

Within the current research, we are trying to develop a unified conceptual and technological approach to support core operation with a complex model by distinguishing concepts and operations on model, data, and system levels. We consider a combination of EC and data-driven approaches as a tool for building intelligent solutions for more precise and systematic managing (and lowering) uncertainty and providing the required level of automation, adaptability, extendibility.

## 2 Conceptual Basis

The proposed approach is based on several key ideas, aimed to extend uncertainty management in complex system modeling and simulation.

1) Disjoint consideration of model, data, and system in terms of structure, behavior, and quality is aimed toward a system-level review of modeling and simulation process and distinguish between the uncertainty of various kinds originated from different level [9].

2) Intelligent technologies like data mining, process mining, machine learning, knowledge-based approaches are to be hired to fill the gap in automation of modeling and simulation. Key sources for the development of such solution include formalization of various knowledge within composite solutions [8] and data-driven technologies to support the identification of model components.

3) EC approaches are widespread in modeling and simulation of complex systems [9,10]. We believe that systematization of this process with separate consideration of spaces for a system (with its sub-systems) and a model (with its sub-models) could enhance such solutions significantly.

4) The aim of the approach's development is twofold. First, it is aimed towards automation of modeling operations to extend the functionality of possible model-based applications. Second, working with a combination of EC and intelligent data-driven technologies could be considered as an additional knowledge source for system and model analysis.

Furthferly this section considers the conceptual basis of the proposed approach with a special focus on the role of EC algorithms and data-driven intelligent technologies for building and exploiting complex models.

### 2.1 Core Concepts

To distinguish between main modeling concepts and operations, we propose a conceptual framework (see Fig. 1) for consideration of key processes and operations during modeling of the complex system. The framework may be considered as a generalization and extension of a framework [11,12] previously defined and used by authors for ensemble-based simulation. Current research extends the concept beyond ensemble-based simulation. It is mainly focused on complex modeling in general with identification of key model management procedures and important artifacts which can be used for model development and application.

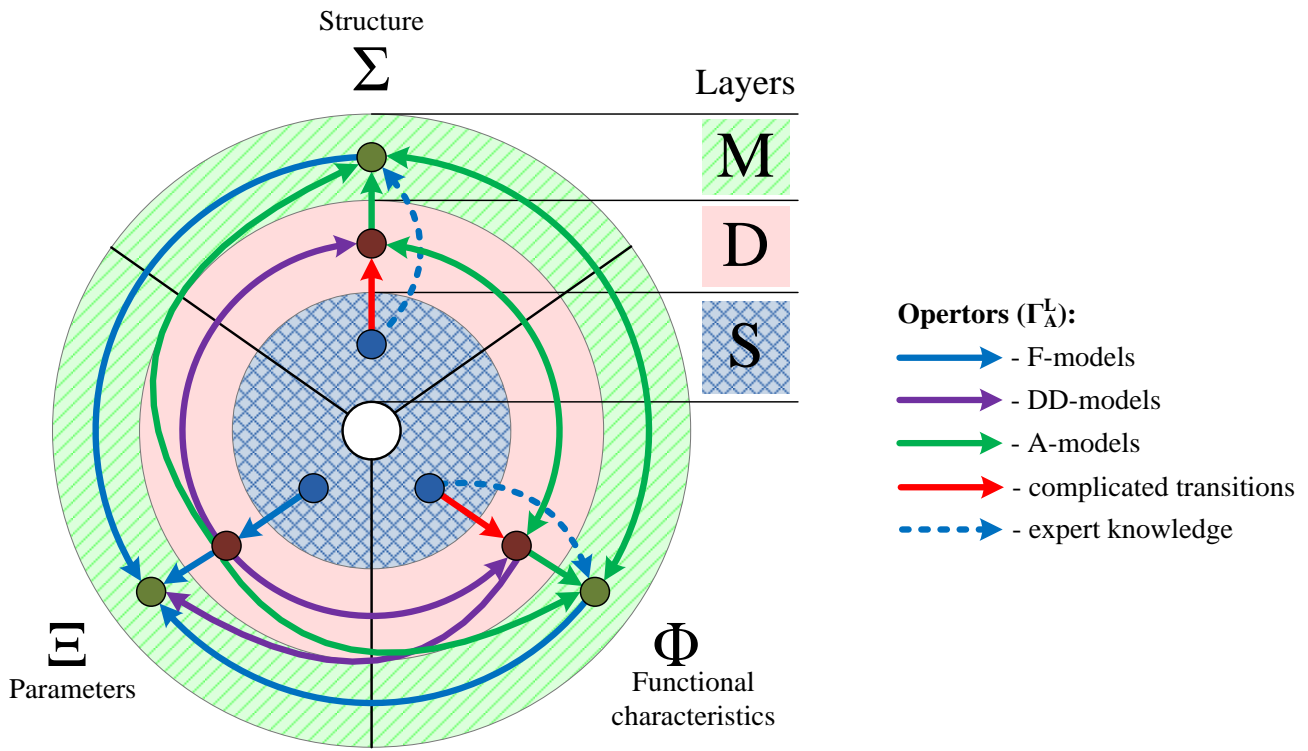

Figure 1 – Basic concepts of complex modeling on model ($M$), data ($D$), and system ($S$) layers

The proposed framework considers three main layers of complex systems' modeling. Namely, model ($M$), data ($D$), and system ($S$). Main operations (arrows on the diagram) within the framework are defined within three concepts: quantitative parameters ($\Xi$), functional characteristics ($\Phi$), and structure ($\Sigma$). We denote operations by $\Gamma_L^A$, where $A$ and $L$ stays for concepts and layer (respectively) involved in the operation. Transitions between concepts and between layers are denoted with $A_1 \rightarrow A_2$ and $L_1 \rightarrow L_2$ respectively, e.g., operator $\Gamma_{S \rightarrow D}^{\Xi}$ reflects observation of quantitative parameters, operator $\Gamma_{D \rightarrow M}^{\Xi}$ stays for basic data assimilation. Also, a set of operators may refer to a single modeling operation, e.g., operators $\Gamma_{M}^{\Phi \rightarrow \Xi}$ and $\Gamma_{M}^{\Sigma \rightarrow \Xi}$ are often implemented within a single monolithic model. Mainly, operators are related to the specific sub-model within a complex model. We consider three key classes of models. F-models are usually classical continuous models developed with knowledge of a system. DD-models are data-driven models based on analysis of available data sets with corresponding techniques (statistics, data mining, process mining, etc.). A-models are mainly intelligent components of a system usually based on machine learning or knowledge-based approaches. Also, we consider EC-based components as belonging to A-models class.

A key problem within complex system modeling and simulation is related to the absent or at least significantly limited possibility to observe the structure and functional characteristics of the system (operators $\Gamma_{S \rightarrow D}^{\Phi}$ and $\Gamma_{S \rightarrow D}^{\Sigma}$) directly. The general solution usually includes implicit substitution of the operators with the expertise of modeler (operators $\Gamma_{S \rightarrow M}^{\Phi}$ and $\Gamma_{S \rightarrow M}^{\Sigma}$). Still, the more complex the system under investigation and the model are, the more limited are those operations. To overcome this issue, additional DD-models are involved (operators $\Gamma_{D}^{\Xi \rightarrow \Sigma}$ and $\Gamma_{D}^{\Xi \rightarrow \Phi}$ for mining in available data, $\Gamma_{D \rightarrow M}^{\Phi \rightarrow \Xi}$ for extended discovery of model parameters for various functional characteristics). Also, A-models are hired to extend expert knowledge in discovery of $M$-layer concepts with either formalized knowledge, or knowledge discovered in data with machine learning approaches (operators $\Gamma_{D \rightarrow M}^{\Xi \rightarrow \Sigma}$, $\Gamma_{D \rightarrow M}^{\Xi \rightarrow \Phi}$, $\Gamma_{D \rightarrow M}^{\Sigma}$, $\Gamma_{D \rightarrow M}^{\Phi}$ for direct discovery of structure and functional characteristics directly and operators $\Gamma_{D}^{\Sigma \rightarrow \Phi}$, $\Gamma_{D}^{\Phi \rightarrow \Sigma}$, $\Gamma_{M}^{\Sigma \rightarrow \Phi}$, $\Gamma_{M}^{\Phi \rightarrow \Sigma}$ for interconnection of discovered characteristics in available data and within the used model). In the proposed approach, primary attention is paid to these kinds of solutions where DD- and A-models enable enhancement of complex modeling process with an additional level of automation, adaptation, and knowledge providing.

**2.2 Complex Modeling Patterns**

Considering the defined conceptual framework, we identify several patterns of modeling and simulation of a complex system (see Fig. 2). The patterns are defined as combinations in a context of the framework described previously (3 layers, 3 concepts). An essential idea of the proposed patterns is systematization of complex model management approaches with combinations of expertise, intelligent solution (A-models), DD-models, and EC.

The pattern extends the operators described in Section 2.1 for model building with operators for model application (i.e. modelling and simulation) and results analysis (e.g. assessing model quality) required for automated model identification. These additional Operators are denoted with $\Gamma'^{A}_{L}$ and similar notation for indices.

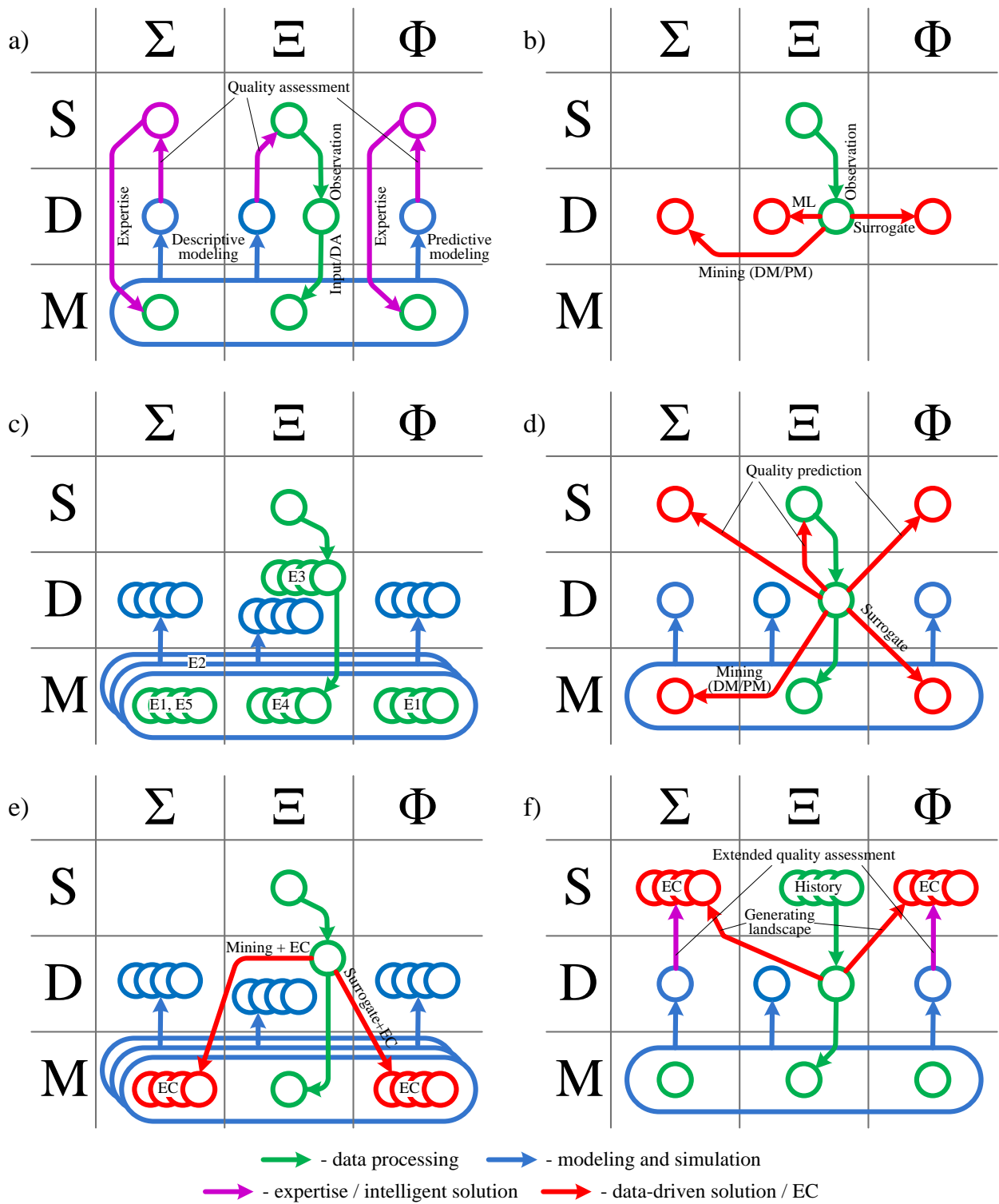

Figure 2 – Complex modeling patterns: a) regular modeling; b) data-driven modeling; c) ensemble-based modeling; d) data-driven support of complex modeling; e) EC in hybrid complex modeling; d) evolutionary space discovery in hybrid complex modeling

*P1.* Regular modeling of a system (Fig. 2a) is a basic pattern usually applied to discover new knowledge on the system under investigation. A model is built using a) expertise of modeled for identification of structure and functional characteristics of the model ($\Gamma_{S \to M}^{\Phi}$ and $\Gamma_{S \to M}^{\Sigma}$); b) available input data usually representing quantitative parameters of a system considered as a static input of the model or source for data assimilation (DA) via operators $\Gamma_{S \to D}^{\Xi}$ and $\Gamma_{D \to M}^{\Xi}$. Results of model application ($\Gamma'^{\Xi}_{M \to D}$, $\Gamma'^{\Sigma}_{M \to D}$, $\Gamma'^{\Phi}_{M \to D}$) could be considered from descriptive (mainly structural or quantitative characteristics) or predictive (often forecasting or other functional characteristics). The obtained results are analyzed in comparison to available information about the investigated system ($\Gamma'^{\Xi}_{D \to S}$, $\Gamma'^{\Sigma}_{D \to S}$, $\Gamma'^{\Phi}_{D \to S}$) forming an optimization loop which can be considered within the scope of all three concepts. Certain limitations within this pattern being applied to complex system modeling and simulation are introduced by two factors. First, working with complex structural and functional characteristics of the model requires a high level of expertise which leads to a limitation of extensibility and automation of model operation. Second, performing optimization in a loop with most algorithms require multiple runs of a model. As a result, computational-intensive models have limitations in optimization-based operations (identification, calibration, etc.) due to performance reasons.

*P2.* Data-driven modeling (Fig. 2b) provides an extension to the modeling operation describing the relationship between data attributes. Application of data-driven models may be considered as replacement of actual "full" model, providing a) information about structure of system and model with data mining (DM) and process mining (PM) techniques ($\Gamma_{D}^{\Xi \to \Sigma}$); b) generating surrogate models for functional characteristics ($\Gamma_{D}^{\Xi \to \Phi}$); c) providing estimation of investigated parameters with machine learning (ML) algorithms and models ($\Gamma'^{\Xi \to \Xi}_{D}$). In contrast to the previous pattern data-driven models usually operate quickly (although it could require significant time to train the model). Still, such models have lower quality than original "full" models. Nevertheless, combining this pattern with others provide significant enhancement in functionality and performance, e.g., data-driven models can be used in optimization loop (see previous pattern).

*P3.* Ensemble-based modeling (Fig. 2c) extends P1 for working with sets of objects (models, data sets, states) reflecting uncertainty, variability, or alternative solutions (e.g., models). Previously [11] we identified 5 classes of ensembles (see E1-E5 in Fig. 2c): decomposition ensemble, alternative models ensemble, data-driven ensemble, parameter diversity ensemble, and meta-ensemble. All these patterns can be applied within a context of the proposed framework. Still, an extension of ensemble structure increases structural complexity of the model and thus leads to the need for additional (automatic) control procedures. Moreover, the performance issues of P1 are getting even worthier in ensemble modeling.

*P4.* One of the key ideas of the proposed approach is an implementation of data-driven analysis of model states, structure, and behavior. To implement it within a conceptual framework we propose pattern for data-driven complex modeling (Fig. 2d). It includes identification and prediction of a model structure through DM and PM techniques ($\Gamma_{D \to M}^{\Xi \to \Sigma}$) and generation of surrogate models for injection into the complex model ($\Gamma_{D \to M}^{\Xi \to \Phi}$). In addition, it is possible to use data-driven techniques to predict the quality of the considered model and use it for model optimization ($\Gamma'^{\Xi}_{D \to S}$, $\Gamma'^{\Xi \to \Sigma}_{D \to S}$, $\Gamma'^{\Xi \to \Phi}_{D \to S}$).

*P5.* A key pattern for EC implementation is presented in Fig. 2e. Here EC is used to identify a model structure ($\Gamma_{D \to M}^{\Xi \to \Sigma}$) and surrogate sub-models ($\Gamma_{D \to M}^{\Xi \to \Phi}$) with a consideration of population of models. As a result, modeling result is also (as well as in P3) presented in multiple instances which may be analyzed, filtered end evolved within consequent iterations over changing time (and processing of coming observations of the system) or within a single timestamp (and fixed observation data).

*P6.* Finally, last presented pattern (Fig. 2f) is aimed at investigation of system phase space using DD-models and/or EC to reflect unobservable landscape for estimation of model positioning, assessing its quality in inferring of (sub-)optimal structural ($\Gamma'^{\Xi \to \Sigma}_{D \to S}$ and $\Gamma'^{\Sigma}_{M \to D}$) and functional ($\Gamma'^{\Xi \to \Phi}_{D \to S}$ and $\Gamma'^{\Phi}_{M \to D}$) characteristics of the actual system.

These patterns could be easily combined to obtain better results within a specific application. Especial interest from the point of view of EC is attracted to the patterns where a set of models (or sub-model) instances is considered (P5, P6). It is possible to consider ensemble-based techniques (P3) in a fashion of EC, but within our approach, we prefer consideration of ensemble as a composite model with several sub-models. In that case, ensemble management refers to the concept of complex model structure.

Several important goals may be reached within the presented patterns:
- automation of complex model management with intelligent solutions, DD-models and EC;
- optimization of model structure and application under defined limitations in precision and performance;
- enhanced ways of domain knowledge discovery for applications and general investigation of a system.

### 2.3 Composite solution development

The proposed structure of core concepts and patterns may be applied in various ways to form a solution which combine operators with original implementation within the solutions or implemented as external model calls. Fig. 3 shows the essential elements (artifacts and procedures) in a typical composite solution within the proposed conceptual layers ($S, D, M$). $S$-layer includes actual system's state which can be assessed through the observation procedure and described by explicit domain knowledge. $D$-layer includes datasets divided into observation data and simulation/modeling data with procedures for data processing and data assimilation. Finally, $M$-layer includes a set of available basic models $M_1 \ldots M_N$ which may be identified, calibrated with available data having tuned models $M'_1 \dots M'_N$ as a result. Here, essential elements are model composition (which may be performed either automatically, or by the modeler) and application of the model.

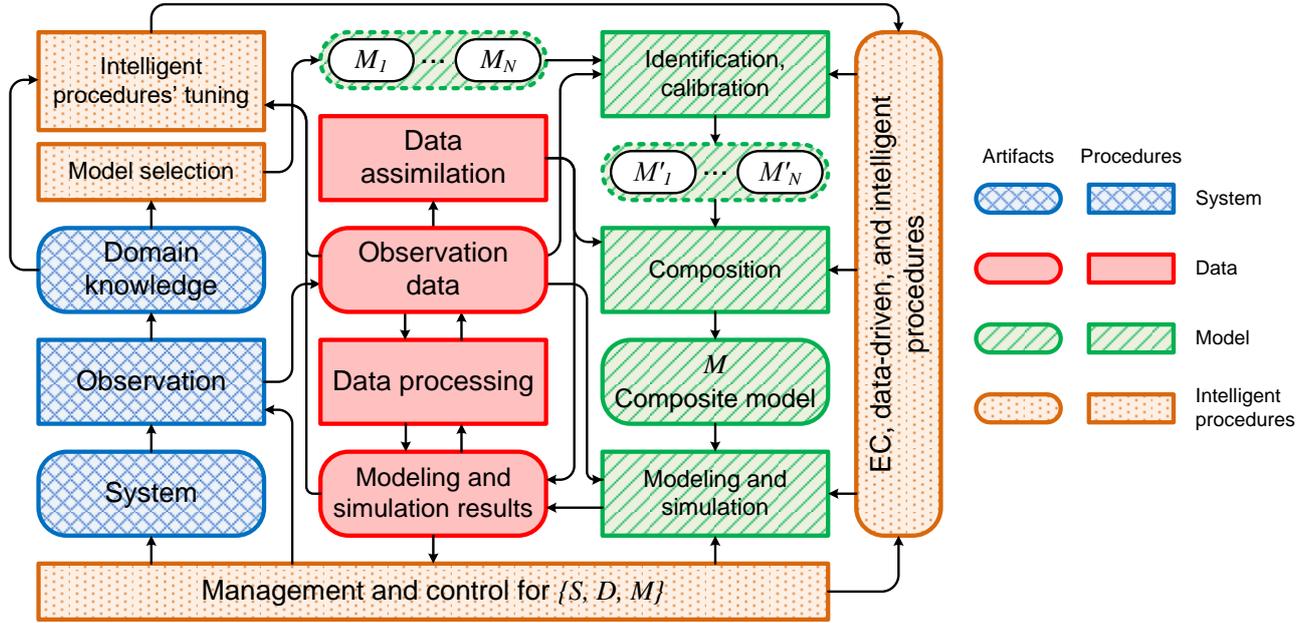

Figure 3 – Artifacts and procedures within a typical composite solution

The key benefit of the approach is an application of a combination of EC, data-driven and intelligent procedures to manage the whole composite solution including data processing, modeling, and simulation to lower uncertainty in $\Sigma \times \Xi \times \Phi$. Within the shown structure these procedures may be applied:

- to rank and select alternative models;
- to support model identification, calibration, composition, and application;
- to manage artifacts on various conceptual layers in a systematic way;
- to infer implicit knowledge from available data and explicitly presented domain knowledge.

The shown example draws a brief view on the composite solution development while the particular details may differ depending on a particular application. Key important procedures within the proposed composite solution are the implementation of intelligent procedures to support model identification and systematic management of composite model are considered in the Sections 2.4-2.5.

**2.4 Evolutionary Model Identification**

Implementing evolution of models within a complex modeling task structure, functional and quantitative parameters are usually considered as genotype whereas model output (data layer) are considered as phenotype. Within the proposed approach we can adapt basic EC operations definition within genotype-phenotype mapping [13]:

- epigenesis as model application: $f_1: S \times M \to D$;
- selection: $f_2: S \times D \to D$;

- genotype survival: $f_3: S \times D \to M$;
- mutation: $f_4: M \to M$.

In addition, we consider quality assessment usually treated as fitness for selection and survival (or, in more complex algorithms for controlling of other operations like mutation):

- data quality: $q_d: S \times D \to Q_d$;
- model quality: $q_m: S \times M \to Q_m$.

Here $Q_d$ and $Q_m$ are often considered as $\mathbb{R}^N$ with some quantitative quality metrics. Model quality usually are considered through data quality, i.e. $q_m \sim q_d(s, f_1(s, m))$, but within our approach this separation is considered as important because in addition we introduce supporting operations with data-driven procedures as in complex modeling many of these functions (first of all $f_1$, $q_m$, $q_d$) have significant difficulties to be applied directly (some of these issues are considered in relationship with patterns). Data driven operations (first of all, $f_1$ and $q_m$) can be introduced as substitution of previously introduced basic operations (see also patterns P2, P4, P6):

- epigenesis as DD-model application: $f_1^d: S \times M \to D$;
- model generation: $g^d: S \times D \to M$;
- model quality prediction: $q_m^d: S \times M \to Q_m$;
- space discovery: $w^d: S \times D \to S$.

Operation $w^d$ could be used within an intelligent extension within selection or survival operations ($f_2$ and $f_3$). It becomes especially important in case of lack of knowledge in system's structure or functional characteristics. Operation $g^d$ at the same time could be used as a part of mutation operation $f_4$ (or initial population generation). Having this extension, we can implement enhanced versions of EC algorithms (e.g., genetic algorithms, evolution strategies, evolutionary programming, etc.) with data-driven operations to overcome or, at least to lower complex modeling issues.

### 2.5 Model Management Approach and Algorithm

By model management we assume operations with models within problem domain solution development and application. This includes identification, calibration, DA, optimization, prediction, forecasting, etc. To systematize the model management in the presented patterns we propose an approach for explicit consideration of spaces $S$, $D$, $M$ within hybrid modeling with EC and DD-modeling. To summarize complex modeling procedures within the approach, we developed a high-level algorithm which includes series of steps to be implemented within a context of complex model management.

***Step 1. Space discovery.*** This step identifies the description of phase space (in most cases, $S$) in case of lack of knowledge or for automation purposes. For example, the step could be applied in the discovery of system state space or model structure. Space description may include a) distance metrics; b) proximity structure (e.g., graph, clustering hierarchy, density, etc.); c) positioning function. One of the possible ways to perform this step is an application of DM and EC algorithm to available data (see pattern P6).

*Step 2. Identification of supplementary functions.* Data-driven functions ($\Phi$) are applied to work in model evolution with consideration of space (landscape) representation as available information.

*Step 3. Evolutionary processing of a set of models.* This step is described by a combination of basic EC operations (population initialization, epigenesis, selection, mutation, survival) with supplementary functions. A form of combination depends on a) selected EC algorithm; b) application requirements and restrictions; c) model-based issues (e.g., performance, quality of surrogate models, etc.).

*Step 4. Assimilation of updated data and knowledge.* This step is applied for automatic adaptation purposes and implement DA algorithm. DA can be applied to a) set of models, b) EC operations (e.g., affecting selection function); c) supplementary functions (as they are mainly data-driven); d) phase space description (if descriptive structure is identified from changed data or/and knowledge).

The steps can be repeated in various combination depending on an application and implemented pattern. Also, the steps are general and could be implemented in various ways. Several examples are provided in the Section 3.

**2.6 Available building blocks of a composite solutions**

EC proposes a flexible and robust solution to identify complex model structures within a complex landscape with possible adaptation towards changing condition and system's state (including new states without prior observation. A significant additional benefit is an ability to manage alternative solutions simultaneously with possible switching and various combination of them depending on the current needs. Still, within the task of model identification and management, the EC (and also many meta-heuristics) have certain drawbacks which require additional steps to implement the approach within particular conditions:

- high computational cost due to the multiple runs of a model;
- low reproducibility and interpretability of obtained results due to randomized nature of the searching procedure;
- complicated tuning of hyper-parameters for better EC convergence;
- indistinct definition of genotype boundaries;
- complicated mapping of genotype to phenotype space.

To overcome these issues, the proposed approach involves two options. First, the intelligent procedures may be used to tune EC hyper-parameters (P5), predict features of genotype-phenotype mapping, boundaries etc. (P4), discover interpretable states and filters (for system, data, and model) to control convergence and adaptation of population (P2, P4, P5) with interpretable and reproducible (through the defined control procedure). Second, the composite model may use various approaches, methods, and elements to obtain better quality and performance of the solution:

- surrogate models (P2, P4, P5) which may increase performance (for example within preliminary and intermediate optimization steps);
- ensemble models (P3) which may be considered as interpretable and controllable population;
- interpretation and formal inference using explicit domain-specific knowledge and results of data mining to feed procedures of EC and infer parameters in both models and EC.
- controllable space decomposition (P6) with predictive models for possible areas and directions of population migration in EC to explicitly lower uncertainty and obtain additional interpretability;

Finally, an essential feature of the proposed approach is a holistic analysis of a composite solution with possible co-evolution models (sub-modes within a composite model) and data processing procedures.

## 3 Application Examples

This section presents several practical examples where the proposed approach, patterns, or some of their elements were applied. The examples were intentionally selected from diverse problem domains to consider generality of the approach. The considered problems are developed in separated projects which are in various stages. Problem #1 (ensemble metocean simulation) were investigated in a series of projects (see, e.g. [11,14,15]). Within this research we are trying to extend model calibration and DA with EC techniques to develop more flexible and accurate multi-model ensembles. Problem #2 (clinical pathways (CPs) modelling) is important in several ongoing project aimed to model-based decision support in healthcare (see, e.g. [16–18]). The proposed approach plays important role by enabling deeper analysis of clinical pathways in various scenarios (interactive analysis of available CPs with identification of clusters of similar patients, DA in predictive modelling of ongoing cases etc.). Finally, Problem #3 shows very early results in recently started project in online social network analysis.

### 3.1 Problem #1: Evolution in Models for Metocean Simulation

The environmental simulation systems usually contain several numerical models serving for different purposes (complementary simulation processes, improving the reliability of a system by performing alternative results, etc.). Each model typically can be described by a large number of quantitative parameters and functional characteristics that should be adjusted by an expert or using intelligent automatized methods (e.g., EC). Alternative models inside the environmental simulation system can be joined in ensemble according to complex modeling pattern based on evolutionary computing (a combination of P3 and P5 patterns). In the current case study, we introduce an example illustrated an ensemble concept in forms of the alternative models ensemble, parameter diversity ensemble, and meta-ensemble. For identification of parameters of proposed ensembles (in a case of model linearity) least square method or (in a case of non-linearity) optimization methods can be used. As we need to take into account not only functional space $\Phi$ and space of parameters $\Xi$ for a single model but also perform optimal coexistence of models in the system (i.e. $\Sigma$), evolutionary and co-evolutionary approaches seem to be an applicable technique for this task. It is worth to mention that co-evolutionary approach can be applied to independent model realizations through an ensemble as a connection element. In this case parameters (weights) in the ensemble can be estimated separately from the co-evolution procedure in a constant form or dynamically. As a case study of complex environmental modeling we design ensemble model that consists of the SWAN[1] model for ocean wave simulation based on two different surface forcings by NCEP[2] and ERA Interim[3]. Thus, different implementations of SWAN model were connected in the form of an alternative models ensemble with least-squares-calculated coefficients defining structure of the complex model. Two parameters – wind drag and whitecapping rate (WCR) – was calibrated using evolutionary and co-evolutionary algorithms implementing $\Gamma_{D \to M}^{\Xi \to \Phi}$ in P5 (for detailed sensitive analysis of SWAN see [19]). Case of co-evolutionary approach can be represented in a form of parameter diversity ensemble, where each population is constructed an ensemble of alternative model results with different parameters. Also, we can add ensemble weights to model parameters diversity and get meta-ensemble that can be identified in a frame of co-evolutionary approach.

In a process of model identification and verification, measurements from several wave stations in Kara sea were used.. Fitness function represents the mean error (RMSE) for all wave stations. For results verification MAE (mean absolute error) and DTW (dynamic time wrapping) metrics were used.

---

[1] http://swanmodel.sourceforge.net/
[2] https://www.esrl.noaa.gov/psd/data/gridded/data.ncep.reanalysis.html
[3] http://apps.ecmwf.int/datasets/data/interim-full-dailys

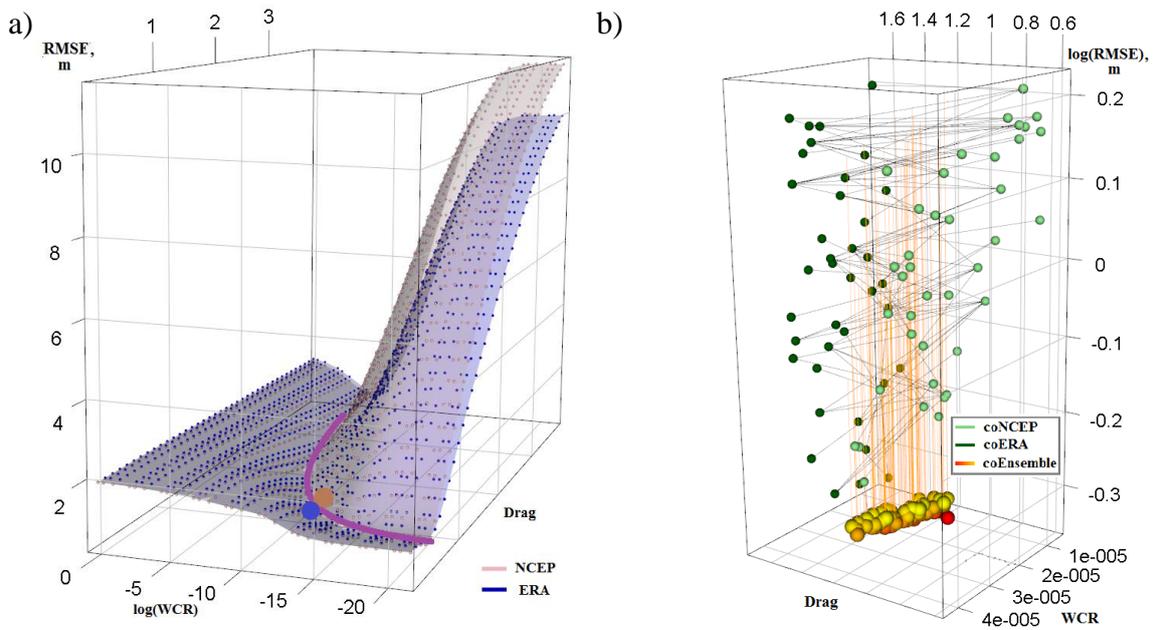

Figure 4 – Metocean simulation a) error landscape for wave height simulation results using ERA and NCEP reanalysis as input data b) Pareto frontier of co-evolution results for all generations

Fig. 4a represents surface (landscape) of RMSE in the space of announced parameters (drag and WCR) for implementations SWAN+ERA and SWAN+NCEP. It can be seen that the evolutionary-obtained results converge to the minimum of possible error landscape. The landscape was obtained by starting the model with all parameters variants from full 30x30 grid (i.e., 900 runs), while evolutionary algorithm was converged in 5 generations with 10 individuals (parameters set) in population (50 runs), that allows performing identification two orders faster. The convergence of co-evolution for SWAN+ERA+NCEP case is presented in Fig. 5.

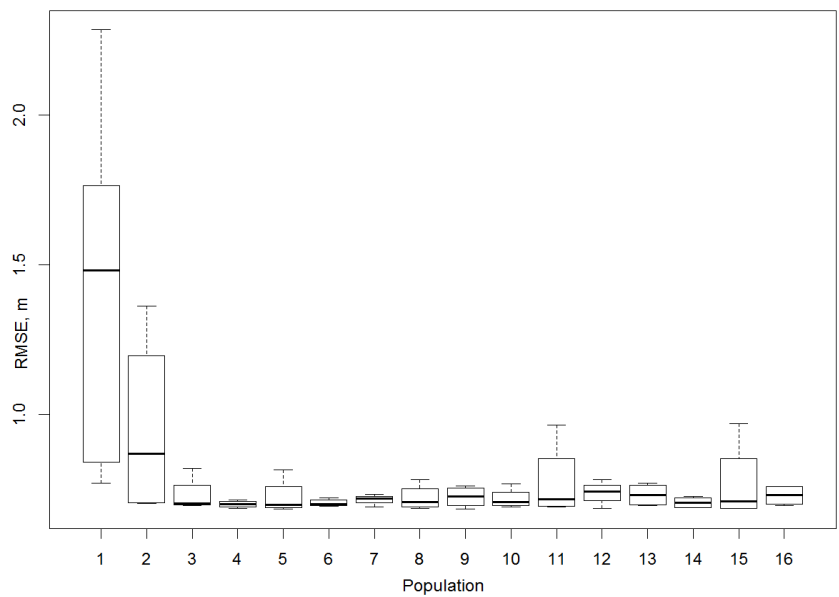

Figure 5 – Co-evolution convergence of diversity parameters ensemble for metocean models

Although error landscapes for a pair of implementations SWAN+ERA and SWAN+NCEP are close to each other, separated evolution does not consider optimization of ensemble result. For this purpose, we apply co-evolutionary approach that produces the set of Pareto-optimal solutions for each generation. Fig. 4b shows that the error of each model in the ensemble is significant (coNCEP and coERA for models along), but the error of the whole ensemble (coEnsemble) converges to minimum very fast.

Obtained result can be analyzed from the uncertainty reduction point of view. Model parameters optimization helps to reduce parameters uncertainty that can be estimated through error function. But when we apply an ensemble approach to evolutionary optimized results, it is suitable to talk about reduction of the uncertainty connected with input data sources (NCEP and ERA) as well. Moreover, meta-ensemble approach allowed to reduce uncertainty, connected with ensemble parameters.

Summarizing results of the metocean case study we can denote that EC approach shows significant efficiency up to 120 times compared with grid search without accuracy losses. According to this experimental study, quality of ensemble with evolutionary optimized models is similar to results of the grid search, MAE metric is equal to 0.24 m and DTW metric – 51. Also, we can mention that co-evolutionary approach provides 10 % accuracy gain compared with results of single evolution of model implementations, but this is still similar to ensemble result with evolutionary optimized models. Nevertheless, co-evolutionary approach allowed to achieve 200 times acceleration. Within the context of the proposed approach space $\Phi$ were investigated using defined structure of the model in space $\Sigma$ for the purpose of model calibration.

### 3.2 Problem #2: Modeling Health Care Process

Modeling health care processes are usually related to the enormous uncertainty and variability even when modeling single disease. One of the way to identify a model of such process is PM [20]. Still, direct implementation of PM methods does not remove a major part of the uncertainty. Within current research, we applied the proposed approach for identification purposes both in the analysis of historical cases and prediction of single process development. Here we consider processes of providing health care in acute coronary syndrome (ACS) cases which is usually considered as one of the major death causes in the world. We used a set of 3434 ACS cases collected during 2010-2015 in Almazov National Medical Research Centre[4] one of the leading cardiological centers in Russia. The data set contains electronic health records of these patients with all registered events and characteristics of a patient.

---

[4] http://www.almazovcentre.ru/?lang=en

To simplify consideration of multi-dimensional space of possible processes ($\Gamma'^{\Xi\to\Sigma}_{D\to S}\Gamma^{\Xi}_{S\to D}$ for analysis of $\Sigma$ on layer $S$) we introduced graph-based representation of this space with vertices representing cases and edges representing proximity of cases. Analysis of such structure enable easy discovering of common cases (e.g. as communities in graph). Such discovering enables explicit, interpretable structuring of the space and representation of further landscape for EC in terms of P6 pattern. Moreover, direct interactive investigation of visual representation of such structure (see Fig. 6) provides significant insights for medical researchers.

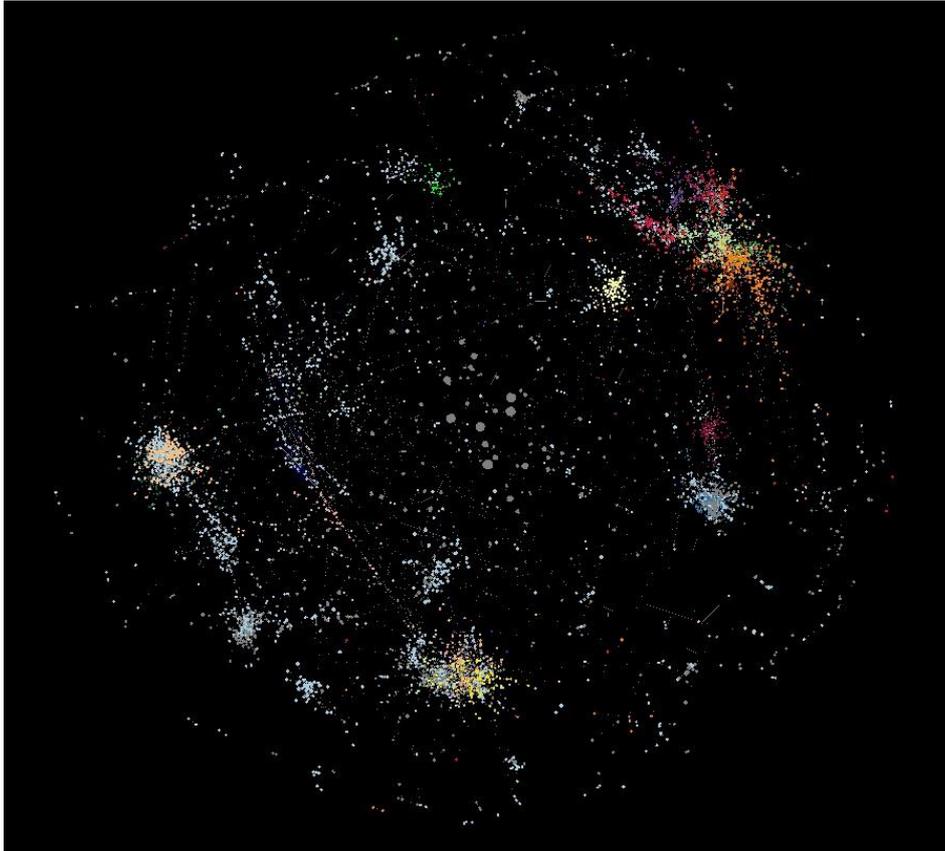

Figure 6 – Graph-based representation of processes space in healthcare (interactive view)[5]

We've developed evolutionary-based algorithm for patterns identification and clustering in such representation with two criteria to be optimized (see Fig. 7). Here processes were represented by a sequence of labels (symbols) denoting key events in PM model. Typical patterns were then selected for Pareto frontier. The convergence process is demonstrated in Fig. 8 (10 best individuals from Pareto frontier according to the integral criterion were selected). As a result, this solution may refer to P5 pattern and operator $\Gamma^{\Xi\to\Sigma}_{D\to M}$ while discovering model structure. Fig. 9 shows an example of typical process model (i.e. structural characteristic of the model) for one of the identified clusters. Detailed description of the approach, algorithms, and results on CPs discovering, clustering and analysis including comparison of three version of CP discovery algorithms with performance comparison can be found in [10]. An important outcome of the approach being applied in this application is interpretability of the clusters and identified patterns. For example, 10 clusters and corresponding CPs obtained interpretation by cardiologists from Almazov National Medical Research Centre. The obtained interpretation and further discovering and application with CP structure are presented in [17]. Another important benefit given by such space structure discovering is lowering uncertainty of patient's treatment trajectory by a hierarchical positioning of an evolved process (selection of a cluster and selection of position within the cluster). For example, discrete-event simulation model described in [17] provides a more appropriate length of stay distribution within simulation with discovered classes of CPs (Kolmogorov-Smirnov statistics decreased by 51% (from 0.255 to 0.124).

---

[5] Demonstration available at https://www.youtube.com/watch?v=EH74f1w6EeY

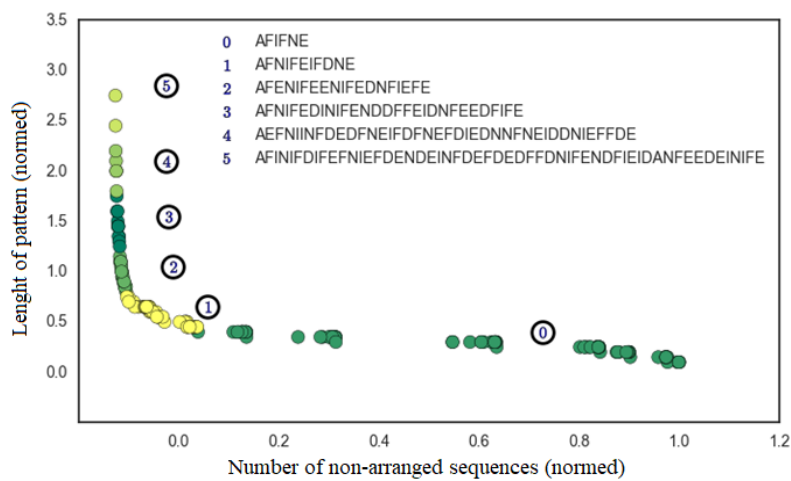

Figure 7 – Pareto frontier for CP patterns discovery

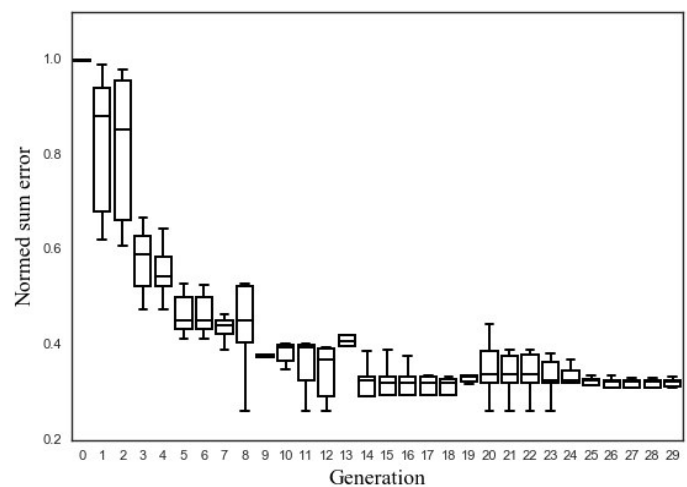

Figure 8 – Evolutionary convergence during CP pattern discovery

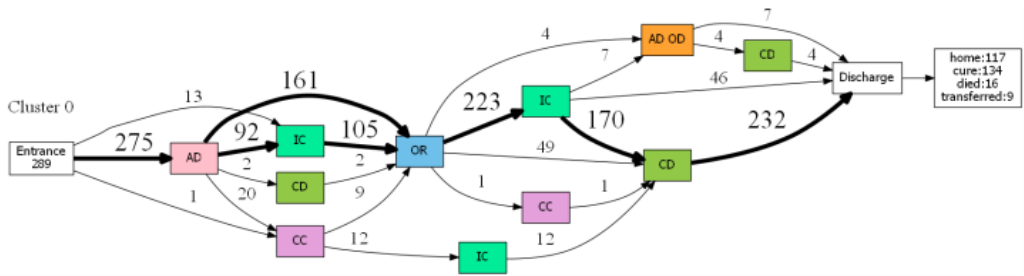

Figure 9 – Example of process model showing transfers between hospital's departments

Furtherly we propose an algorithm to dynamically generate possible development of the process in healthcare using identified graph-based space representation with evolutionary strategies, assimilating incoming data (events) within a case ($\Gamma'^{\Sigma}_{M \to D}$ in P5 and $\Gamma^{\Xi \to \Sigma}_{D}$ in P2). We consider convergence (Fig. 10a) of the introduced synthetic continuation of the processes to the right class (identified clusters of typical cases were used) with mapping to the graph-based space representation with proximity measures (Fig. 10b). As a result, the appearance of the CP's events decreases the number of synthetic CPs and increase percentage of CPs positioned in the correct cluster (see an example in Fig 10c and Fig. 10d correspondingly). This enables interpretable positioning and uncertainty lowering in predicting further CP's development for a particular patient.

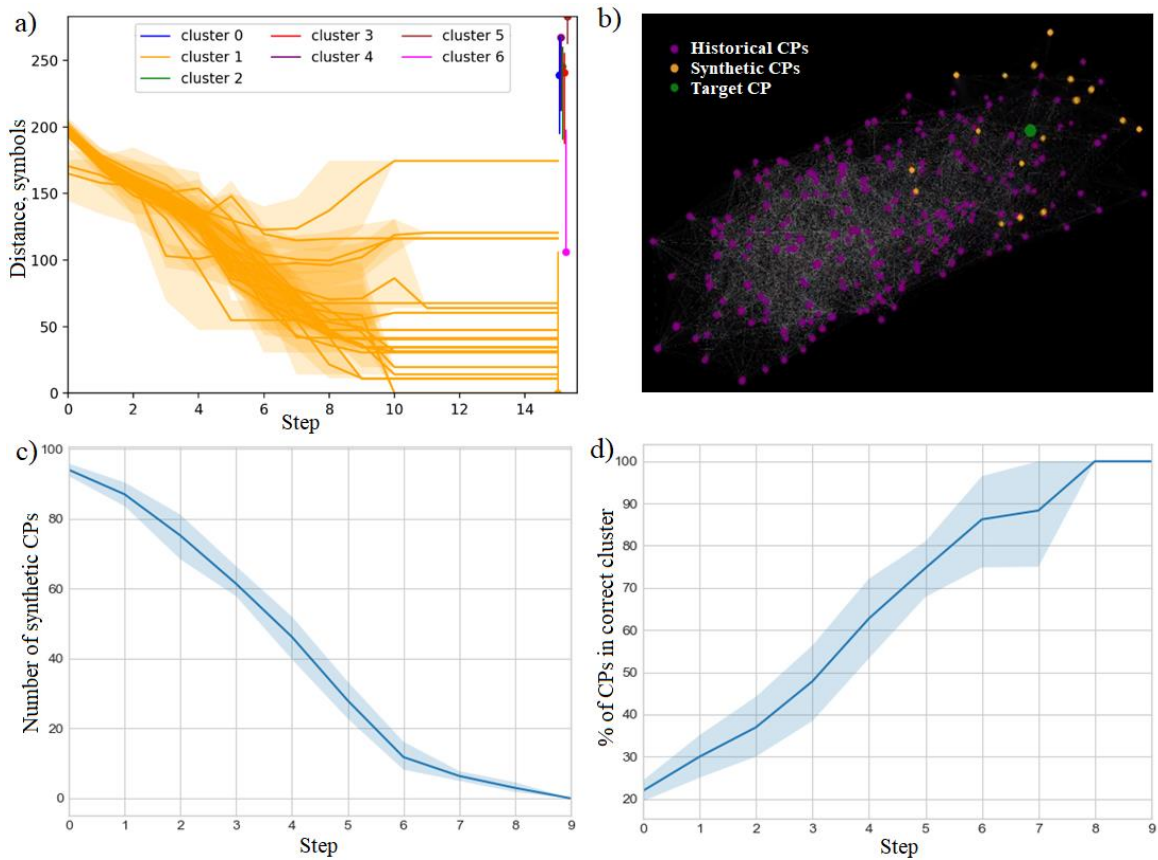

Figure 10 – Evolution of synthetic CPs a) CP population convergence; b) evolution of possible CP[6]; c) number of synthetic CPs; d) % of CPs in correct cluster

---

[6] Demonstration available at https://www.youtube.com/watch?v=twvfX9zKsY8

Here a combination of patterns P2, P5, and P6 in the implementation of the proposed algorithm (see Section 2.5) enable interactive investigation of processes space and data assimilation into a population of possible continuations of a single process during its evolving. This solution can be applied in exploratory modeling and simulation of patient flow processing as well as decision support in specialized medical centers.

### 3.3 Problem #3: Mining Social Media

Nowadays social media analysis (that began with static network models emphasizing a topology of connections between users) strives to explore dynamic behavioral patterns of individuals which can be recovered from their digital traces on the web. The prediction of social media activities requires to combine analytical and data-driven models as well as to identify the optimal structure and parameters of these models according to the available data. Here we show an example of the problem in this field involving evolutionary identification of a model.

A digital trace of a user in an online social network (OSN) is a sequence (chain) of observed activities separated with time gaps. Each OSN supports different types of "hidden" and observable activities. For example, in a largest Russian social network vk.com (further is denoted as VK) a user has a personal page (wall) with three types of activities: post (P) – when a user makes a record by himself; repost (R) – when the user copies the record of another user or community to his or her wall and comment (C) – when the user comments the record on his or her wall. Fig. 11 illustrates the distribution of these activities for subscribers of large Russian bank community in VK. The collected dataset consists of 100 (or less if unavailable) last entries (posts or reposts), and comments for the entries for 8K user walls in a period January 2017–December 2017. Comments are much less common than posts and reposts. The distributions of the posts and reposts are similar, but there is a group of "spreaders" with a significant number of reposts.

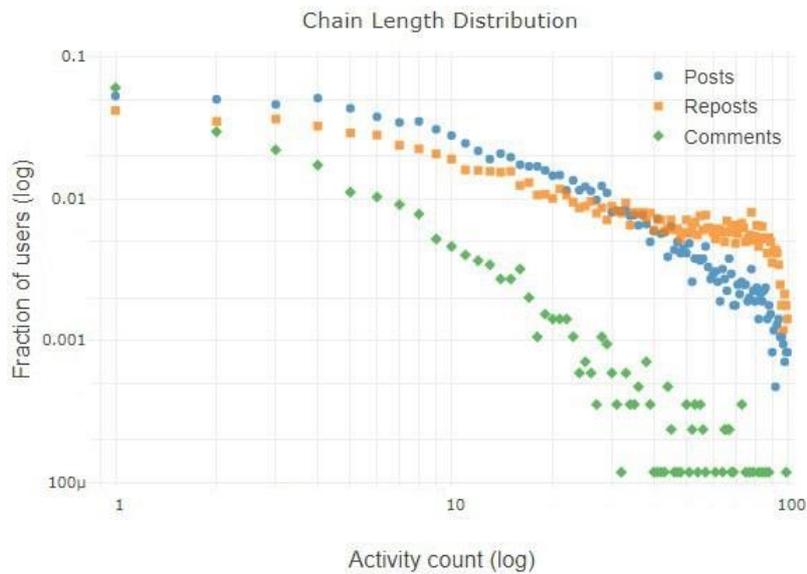

Figure 11 – Distribution of posts, reposts and comments on personal walls of subscribers of bank community

We applied the technique described in Section 3.2 to analyze the processes. Still, the considered process has significantly different structure. By default, it is continuous with random repetition of events, while healthcare process in ACS cases has finite and more "strong" structure. Fig. 12 shows a typical process structure identified with EC-based approach and visualized with expanded cycles (a) and with collapsed cycles (b). The second one could be considered as more relevant than the first one which is significantly affected by a length of selected history. It is natural to consider it as a random process or state-transition model. In that case, three identified clusters (characterized by various frequencies of transitions) could be interpreted as typical behavior models.

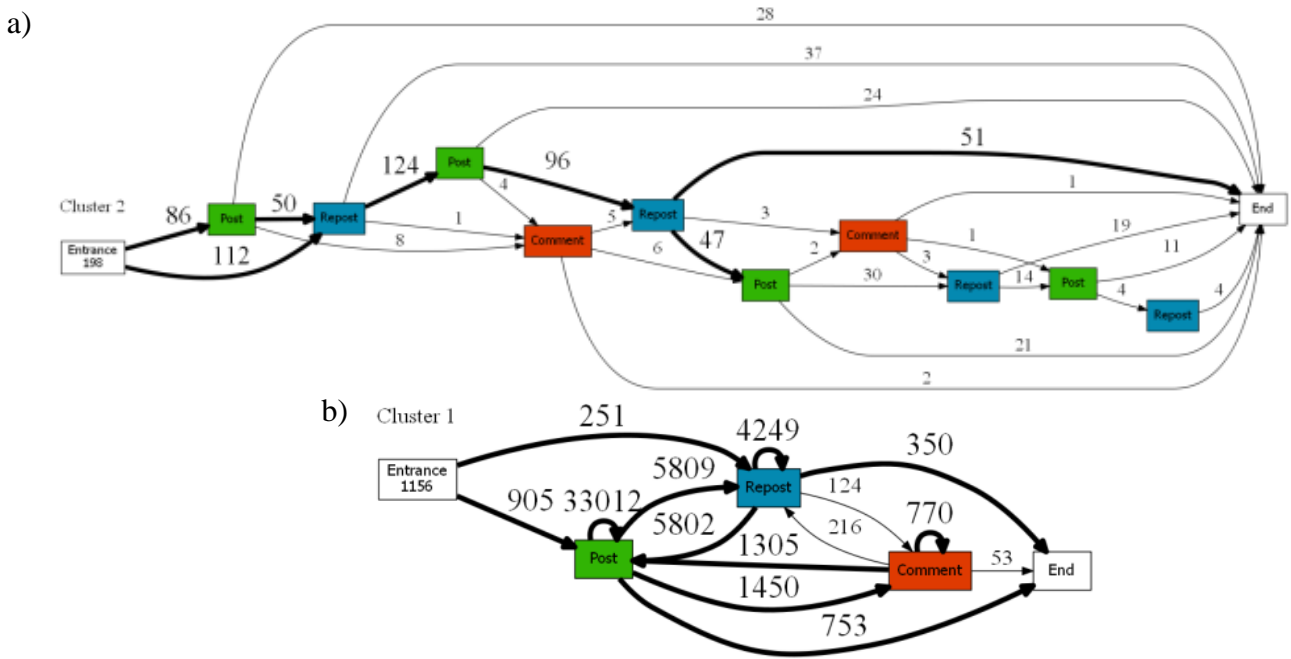

Figure 12 – Example of process model a) with expanded cycles; b) with collapsed cycles

N-grams analysis is often used to detect patterns in people's behaviors [21,22]. N-grams analysis is based on counting frequencies of combinations or sequences of activities. We collected all sorted 3-grams (so called 3-sets) for each user's sequence to analyze the frequency of event combinations. As a result, three clusters of vectors with 3-sets chains were identified with k-means clustering method. Fig. 13 shows all combinations and transitions between them for cluster #3 as an example. Using fig. 12 and table 1, it is possible to see that cluster #3 includes users who often make new records (P) and sometimes comment records (C). So, cluster #3 mostly consists of "bloggers". Cluster #2 includes "spreaders" who copies other records (R) frequently. And the biggest cluster #1 consists of people who make new records and copies other ones equally but less intensive comparing to other clusters. That may be considered as a typical behavior for user of OSN. N-grams analysis allows detecting typical behavioral patterns and obtaining process models for social media activities using chains of different lengths as input data. Thus, this type of data-driven modeling is more appropriate to research continuous processes. Fig. 14 shows a graph-based representation of process space with of all users' patterns.

| Cluster | Size | CCC | CCP | CCR | CPP | CPR | CRR | PPP | PPR | PRR | RRR |
|---|---|---|---|---|---|---|---|---|---|---|---|
| 1 | 5238 | 0.52 | 0.74 | 0.24 | 1 | 0.72 | 0.42 | 6.68 | 6.4 | 7.56 | 8.53 |
| 2 | 2110 | 0.11 | 0.13 | 0.16 | 0.16 | 0.36 | 0.59 | 2.09 | 3.95 | 12.32 | 60.3 |
| 3 | 1120 | 0.91 | 1.38 | 0.14 | 3.62 | 0.75 | 0.18 | 50.02 | 11.78 | 5 | 2.77 |

Table 1 – Mean of activities' combinations for users' clusters

Figure 13 – Example of process model for cluster #3 using n-gram analysis

Figure 14 – Graph-based representation of processes' space
for three clusters in social media activity

This subsection provides very early results. Next step within application of the proposed approach in this application includes an extension of process model structure a) with temporal labeling (gaps between events); b) considering process within a sliding time window to get more structured processes; c) linking the model with causal inference; d) introduction of DM techniques for EC positioning of ongoing processes in model space. We believe that these extensions could enhance discovery of model structure (P4) and provide deeper insight on social media activity investigation.

## 4 Conclusion and Future Work

The development of the proposed approach is still an ongoing project. We are aimed towards further systematization and detailing of the proposed concepts, methods, and algorithms, as well as more comprehensive and deeper implementation of EC-based applications. Further work of the development includes the following directions:

- dualization on the role of data-driven and intelligent operations in proposed approach and described patterns;
- extended analysis of various EC techniques applicable within the approach;
- investigation on EC-based discovery for models of complex systems with lack or inconsistent observations;
- detailed formalization of expertise and knowledge-based methods within the approach;
- extending the approach with interactive user-centered modelling and phase space analysis;
- development of multi-layered approach for decision support and control of system and process $S$, available data $D$, complex model $M$.

*Acknowledgments.* This research is financially supported by The Russian Scientific Foundation, Agreement #14-11-00823 (15.07.2014).